\documentclass[fleqn,usenatbib]{mnras}

\usepackage{newtxtext,newtxmath}

\usepackage[T1]{fontenc}

\DeclareRobustCommand{\VAN}[3]{#2}
\let\VANthebibliography\thebibliography
\def\thebibliography{\DeclareRobustCommand{\VAN}[3]{##3}\VANthebibliography}

\usepackage{graphicx}	
\usepackage{amsmath}	

\title[Accreting MS stars and MT instability]{The radius variations of accreting main sequence stars and mass transfer instability}

\author[Z. Zhao et al.]{
Zi-Qi Zhao,$^{1,2,3}$ 
Zhen-Wei Li,$^{4,5,6}$\thanks{E-mail: lizw@ynao.ac.cn}
Lin Xiao,$^{1,2,3}$
Hong-Wei Ge$^{4,5,6}$
and Zhan-Wen Han$^{4,5,6,7}$
\\
$^{1}$Department of Physics, Hebei University, Baoding, 071002, China\\
$^{2}$Hebei Key Laboratory of High-precision Computation and Application of Quantum Field Theory, Baoding, 071002, China\\
$^{3}$Hebei Research Center of the Basic Discipline for Computational Physics
College of Physics Science and Technology, Hebei University, Baoding 071002, China\\
$^{4}$Yunnan Observatories, Chinese Academy of Sciences, Kunming, 650216, People's Republic of China\\
$^{5}$Key Laboratory for the Structure and Evolution of Celestial Objects, Chinese Academy of Science, People's Republic of China\\
$^{6}$International Centre of Supernovae, Yunnan Key Laboratory, Kunming, 650216, People's Republic of China\\
$^{7}$University of the Chinese Academy of Science, Yuquan Road 19, Shijingshan Block, 100049, Beijing, People's Republic of China\\}
\date{Accepted XXX. Received YYY; in original form ZZZ}

\pubyear{2024}

\begin{document}
\label{firstpage}
\pagerange{\pageref{firstpage}--\pageref{lastpage}}
\maketitle

\begin{abstract}
Many previous works studied the dynamical timescale mass transfer stability criteria based on the donor response with neglecting the stellar structure of the accretor. In this letter, we investigate the radial response of accretors with mass accumulation and its effect on the binary mass transfer stability. We perform a series of detailed stellar evolution simulations with different types of accretors and obtain the radial variations of stars accreting at different rates. Since the time within which the donor loses half of the original mass has a correlation with the donor mass, we approximately obtain the mean mass transfer rate as a function of mass ratio. Assuming that the common envelope (CE) phase occurs if the accretor radius exceeds the outer Roche lobe radius, we obtain the critical mass ratio of dynamically unstable mass transfer. We find the critical mass ratios for donors filling their Roche lobes at the Main Sequence (MS) and Hertzsprung Gap (HG) stages are smaller than that derived from the radial response of the donor in the traditional way. Our results may suggest that the binary is easier to enter into the CE phase for a donor star at the MS or HG stage than previously believed.
\end{abstract}




\section{Introduction}
\label{intro}

Stellar physics is the cornerstone of astrophysics, and about (more than) half of stars are born with a binary companion
\citep{2012Sci...337..444S,2013ARA&A..51..269D,2017yCat..18100061M}. 
Binary interactions add complexity to stellar evolution, resulting in the formation of remarkably unique stars and intriguing observational phenomena \citep{2002MNRAS.335..948C,2001PABei..19..242H,2020RAA....20..161H}, such as double black holes, double neutron stars, and type Ia supernovae (SNe Ia), etc \citep{1992PASP..104..717P,2012ApJ...759...52D,2014LRR....17....3P,2017ApJ...846..170T,2020ApJS..249....9G,2023RAA....23h2001L}. Though enormous progress in binary evolution theory has been made in recent decades, there are two longstanding questions still in debate, which are the binary mass transfer stability and common envelope (CE) process, respectively \citep{2003MNRAS.341..669H,2010NewAR..54...39P,2023arXiv231111454C}. Both physical problems exert crucial effects on the binary products \citep{2013A&ARv..21...59I,2019MNRAS.486.5809P,2020ApJ...899..132G,2023A&A...669A..82L}.   

The response of accretor during the mass transfer phase is significant in determining the outcome of binary evolution \citep{1960ApJ...132..146M,2002ASPC..267..165Y,2014ASPC..490..287N}. However, there is a lack of studies focusing on the accretors. \citet{1976ApJ...205..217F} found that the accretor filling its Roche lobe would lead to both mass and system angular momentum loss. Subsequently, \citet{1977A&A....54..539K} simulated a spherically symmetric star model and studied the radius change under constant mass transfer rates. The simulations revealed that the accretor expands as it accumulates material, accompanied by increased luminosity. The radius will increase until the internal pressure and gravity of the star reach equilibrium again. Then, the star contracts and returns to the main sequence (MS). This finding holds significant importance in comprehending the evolution of accretors in binary systems.

Regarding higher mass transfer rates, \citet{1977PASJ...29..249N} found that accretors have a limited capacity to accumulate material. The rest of the material will form a disk-like structure around the star. \citet{1989ApJ...341..306F} calculated the radius variations of a low-mass ($M=0.75M_{\odot}$) MS star during accretion processes. Due to the different internal structures of the low-mass star and more massive star ($M=2M_{\odot}$), the radius changes during accretion will be different. For example, stars will shrink due to the accretion of the convective envelope and expand for that of the radiative envelope. However, the relationship between the variation of radius and stellar mass under accretion has not been studied in detail.

In this letter, we investigate the radial variations of accretors with mass accumulation and aim to find the mass transfer stability on account of the accretor response. The rest of the paper is structured as follows. We give the model inputs and methods in Section \ref{sect:Model Inputs and Methods} and the results in Section \ref{result}. Finally, the summary and conclusion is addressed in section \ref{sect:conclusion}.

\section{Model Inputs and Methods}
\label{sect:Model Inputs and Methods}

\subsection{ Stellar Evolution Code}
\label{mcd}

The calculations are performed with the state-of-the-art evolution code Modules for Experiments in Stellar Astrophysics (MESA, version 12115, \citealt{2011ApJS..192....3P,2013ApJS..208....4P,2015ApJS..220...15P,2018ApJS..234...34P,2019ApJS..243...10P}).
In this letter, we mainly consider the response of accretor during mass transfer processes, and the calculations are performed in single stellar evolution. We start with models of zero-age main-sequence (ZAMS) stars as accretors, then stars accrete material with different mass transfer rates. The solar metallicity of $Z=0.02$ with a hydrogen fraction of $X=0.7$ is adopted (e.g. \citealt{1971MNRAS.151..351E,2009ARA&A..47..481A}). We use the opacity table of Type II OPAL and the mixing-length parameter of $2$ \citep{1999A&A...346..111L}. We assume that the entropy of the accreting material matches the surface of the accretor (e.g., \citealt{2021ApJ...923..277R,2023A&A...669A..45T}). The initial accretor masses are set to be $0.2M_{\odot}$, $0.5M_{\odot}$, $1M_{\odot}$, $2M_{\odot}$, $5M_{\odot}$ and $10M_{\odot}$, which cover different types of stars from fully convective stars to massive stars. The mass transfer rate is taken as a free parameter, and we stop the simulation when the accretor mass increases to four times the initial mass. 

In this work, we mainly focus on the radius variance of a star with the mass accumulation, and the angular momentum accretion is not considered. The angular momentum accretion mainly exerts two effects on the binary evolution: (1) The mass transfer efficiency. In general, the mass gainer will be spun up due to the mass accretion. There is a commonly adopted model, i.e., mass transfer is conservative until the accretor reaches critical rotation, after which rotationally enhanced mass loss governs the mass transfer efficiency  \citep{2005A&A...435.1013P}. In this model, if there is no effective mechanism to prevent the spin-up of accretors, the accretion efficiency is generally small, typically about $0.1-0.4$, as calculated by \citet{2022A&A...659A..98S}. However, it is still unclear whether the critical rotation can effectively stop the mass accretion. Some works argue that the mass transfer efficiency should be large. For example, \citet{2021ApJ...908...67S} considered three mass transfer models and found that the model of mass transfer efficiency of 0.5 matches the observations of Be binaries better. For convenience, we assume that the accretor can absorb all material without mass loss. Nevertheless, the influence of mass transfer efficiency on our results will be discussed in Section \ref{result5}. (2) The rotation would alter the stellar structure. Though the rotation is important, the mass is undoubtedly the most important factor that affects the stellar’s radius \citep{2011A&A...530A.115B}, in particular for stars with masses less than 10$M_{\odot}$ (The maximum accretor mass in our simulations). 
Above all, the simulations in this work are performed with non-rotation models.

\subsection{Thermal Timescale}
\label{id}
During the MS evolution, a star remains in hydrostatic equilibrium and expands slowly due to the nuclear burning. While the accreted material onto the surface of the star would change the thermal properties of accretors. The thermal property of a star is generally understood via the thermal timescale, which is defined as, 
\begin{equation}\label{eq1}
t_{\rm K} = \frac{GM^2}{RL},
\end{equation}
where ${G}$ is the gravity constant, ${M}$ is the total mass, ${R}$ is the radius and ${L}$ is the  luminosity of the star. Then, we can get the thermal timescale mass transfer rate $\dot{M}_{\rm KH}$ via equation \ref{eq1}, i.e.,
\begin{equation}\label{eq2}
\dot{M}_{\rm KH} = \frac{RL}{GM}.
\end{equation}
If $\dot{M} \leq \dot{M}_{\rm KH}$, the accretor has enough time to adjust itself, so the evolutionary track on the Hertzsprung-Russell (HR) diagram is mainly affected by the stellar mass. However, if $\dot{M} \geq \dot{M}_{\rm KH}$, the stellar radius will increase due to the mass accumulation \citep{1976ApJ...206..509U,2004ApJ...601.1058I}. The thermal timescale mass transfer rate $\dot{M}_{\rm KH}$ becomes bigger as a star accretes matter \citep{1997A&A...327..620S}. Therefore, when $\dot{M}_{\rm KH}$ exceeds $\dot{M}$, the star ceases to expand and reverts to MS. Subsequently, the star will ascend along the MS because of the accretion \citep{1987ApJ...318..794H,2004A&A...419.1057W}. In this letter, we mainly focus on the case of $\dot{M} > \dot{M}_{\rm KH}$.

\section{Results}
\label{result}

\subsection{Accretion with Radiative Envelope}
\label{result1}

\begin{figure}

  \begin{minipage}[t]{0.45\textwidth}
  \centering
   \includegraphics[scale=0.33]{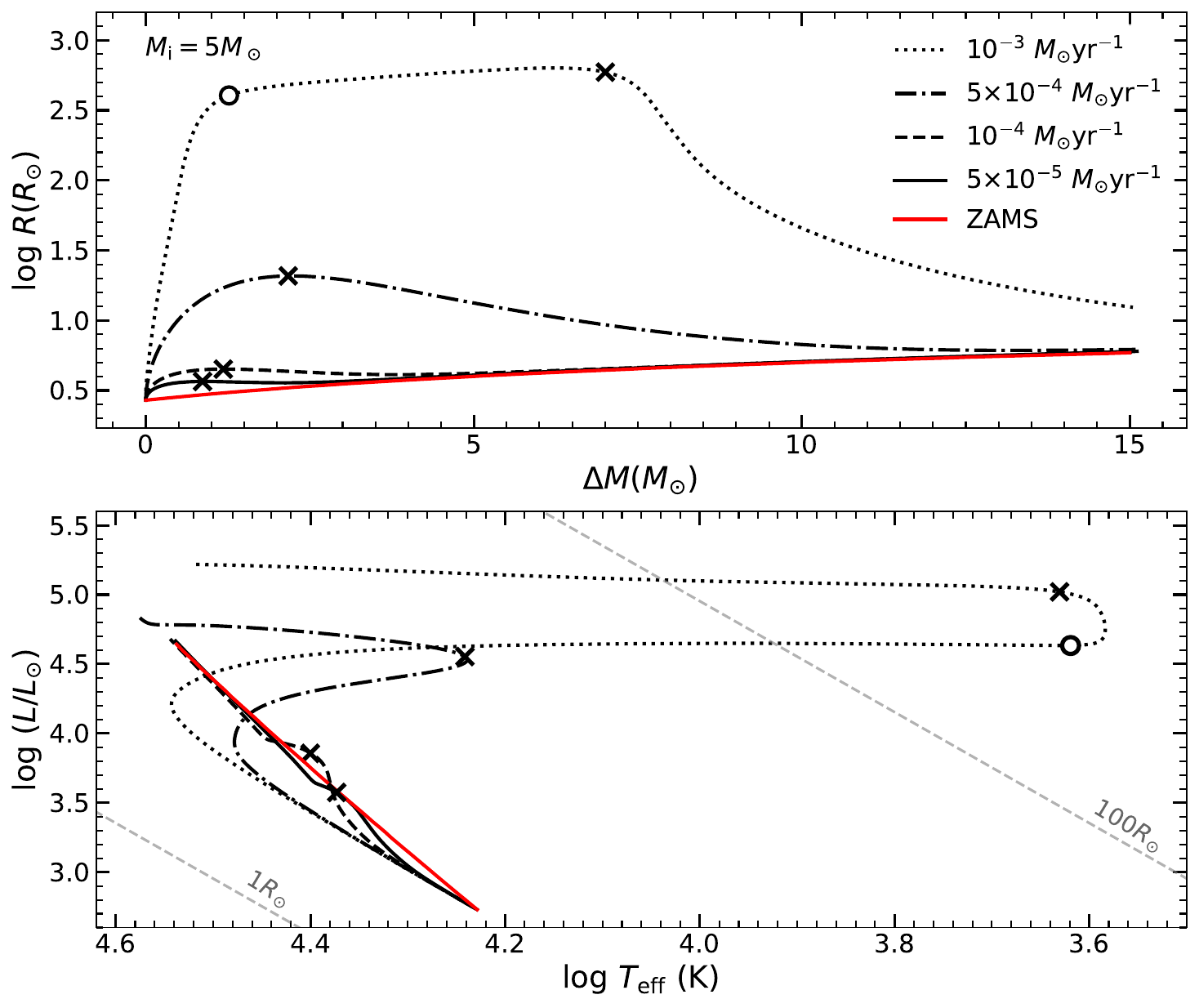}

  \end{minipage}%
   \caption{\label{fig1} \textbf{Upper panel:} The radial variations of accretors with initial masses of $5M_{\odot}$ as a function of increased mass, where $\Delta M= M - M\rm{_i}$ and $M\rm{_i}$ is the star's initial mass. The black solid, dashed, dash-dotted and dotted lines are for different accretion rates, as shown in the legend. 
   The crosses represent the stage of the maximum radius. The open circle represents the appearance of the convective envelope. The red solid line represents the ZAMS radius for accretors with given masses. The thermal timescale mass transfer rate for a $5M_\odot$ accretor is about $1.25\times 10^{-5}M_\odot\;\rm yr^{-1}$. \textbf{Lower panel:} Evolutionary tracks of accretors with initial masses of $5M_{\odot}$.}
\end{figure}

Stars with masses larger than $\sim 1.6\rm{M_{\odot}}$ have convective cores and radiative envelopes \citep{2013sse..book.....K,2020RAA....20..161H}. The existence of a radiative envelope would lead to an increase in envelope thermal energy if mass accumulated onto the surface of the star, resulting in star expansion and evolutionary track deviations from ZAMS \citep{1967AcA....17..355P, 2024arXiv240109570L}.
In the upper panel of Figure \ref{fig1}, we present the radius variation with different mass transfer rates for an initial accretor mass of the $5M_{\odot}$. The HR diagram of the $5M_{\odot}$ accretor is presented in the lower panel of Figure \ref{fig1}. The thermal timescale mass transfer rate is about $1.25{\times}10^{-5} M_{\odot}\; \rm{yr^{-1}}$ for a $5M_{\odot}$ accretor. All of the inputs of mass transfer rates are larger than the corresponding $\dot{M}_{\rm KH}$ for this example. We see that the radius with a given mass transfer rate is always larger than the radius of the ZAMS star (as shown in red solid lines) with the same mass, and the evolutionary tracks move towards lower temperatures relative to ZAMS. 

The radius does not always increase continuously with accretion, as shown in Figure~\ref{fig1}. We note that the radius under different $\dot{M}$ increases at first and then approaches the ZAMS radius, corresponding to the tracks initially moving away from the ZAMS and then returning back. At the beginning of the star accumulating material, the evolutionary tracks of the accretor are located on the left relative to ZAMS in the HR diagram. The reason is that the thermal equilibrium can not be maintained after the star receives a large amount of material. The external energy causes the temperature and luminosity of the star to rise \citep{1977PASJ...29..249N}.This phase is very short, and the star only accretes little matter (e.g., for the case of $\dot{M}$ = $10^{-3} M_{\odot}\; \rm{yr^{-1}}$, $\Delta M= 0.006M_{\odot}$). Therefore, the effect of this process on the radius variance in the upper panel is inconspicuous. The star expands fast in the subsequent evolution, which leads to a decrease in temperature, and the evolutionary track moves right. Then, the radius increases until the maximum radius is reached, represented by crosses in Figure~\ref{fig1}. The corresponding position for stars with maximum radius are also shown in the lower panel. After the star expands to the maximum radius, the given mass transfer rate $\dot{M}$ is smaller than the thermal timescale mass transfer rate $\dot{M}_{\rm KH}$, and the stars can maintain the thermal equilibrium once again, resulting in the fall of radius and the evolutionary tracks return to the neighbor of the ZAMS line. It should be noted that the mass accumulation processes always proceed on thermal timescale or beyond, and nuclear burning is relatively unimportant. Therefore, our evolutionary tracks show the difference in comparison with detailed binary evolution simulations, where the mass transfer rate may proceed on a nuclear timescale (e.g., \citealt{2021ApJ...923..277R,2023ApJ...942L..32R}.)

In Fig~\ref{fig1}, there is an epoch that the radius increases very slowly for the case of $\dot{M}$ = $10^{-3} M_{\odot}\; \rm{yr^{-1}}$, as shown in the part between the circle and cross. The radius variation of this interval is significantly different from the cases with lower $\dot{M}$. We see that the temperature remains almost constant, and the luminosity increases in the corresponding HR diagram of the lower panel. The reason can be understood as follows. For an accretor of $5M_{\odot}$, there is a radiative envelope initially. The surface of the accretor receiving a large amount of energy in a short time (the case of $10^{-3} M_{\odot}\; \rm{yr^{-1}}$) would result in a steep temperature gradient and large entropy. The rapid increase of surface entropy causes the appearance of the surface convective envelope, {which suppresses the stellar radius and temperature. When the star reaches thermal equilibrium again, the convective envelope gradually disappears around the position of maximum radius. In the subsequent evolution, the thermal timescale mass transfer rate $\dot{M}_{\rm KH}$ is larger than the accretion rate, and we see that the radius rapidly decreases back to the ZAMS radius correspondingly.}

Since the expansion behavior and evolutionary track of stars with radii of $2M_{\odot}$ and $10M_{\odot}$ are similar to that of $5M_{\odot}$, they will not be discussed in detail. The results are addressed in Appendix \ref{appendix}.

\subsection{Accretion with Convective Envelope}
\label{result2}

When the stellar mass is less than $\sim 1.6 M_{\odot}$, the star has a structure of convective envelope and internal radiative core \citep{2020RAA....20..161H,2020ApJ...899..132G}. The depth of the convective envelope is increased with the decrease of the stellar mass. Specifically, the star may have a fully convective structure with a mass of less than $\sim 0.35 M_{\odot}$ \citep{2008ApJ...676.1262B}. In this section, we study the cases of accretors with shallow convective envelopes, deep convective envelopes and complete convection.

\subsubsection{Accretor with shallow convective envelope}
\label{result3}

\begin{figure}
  \begin{minipage}[htp]{0.45\linewidth}
  \centering
   \includegraphics[scale=0.33]{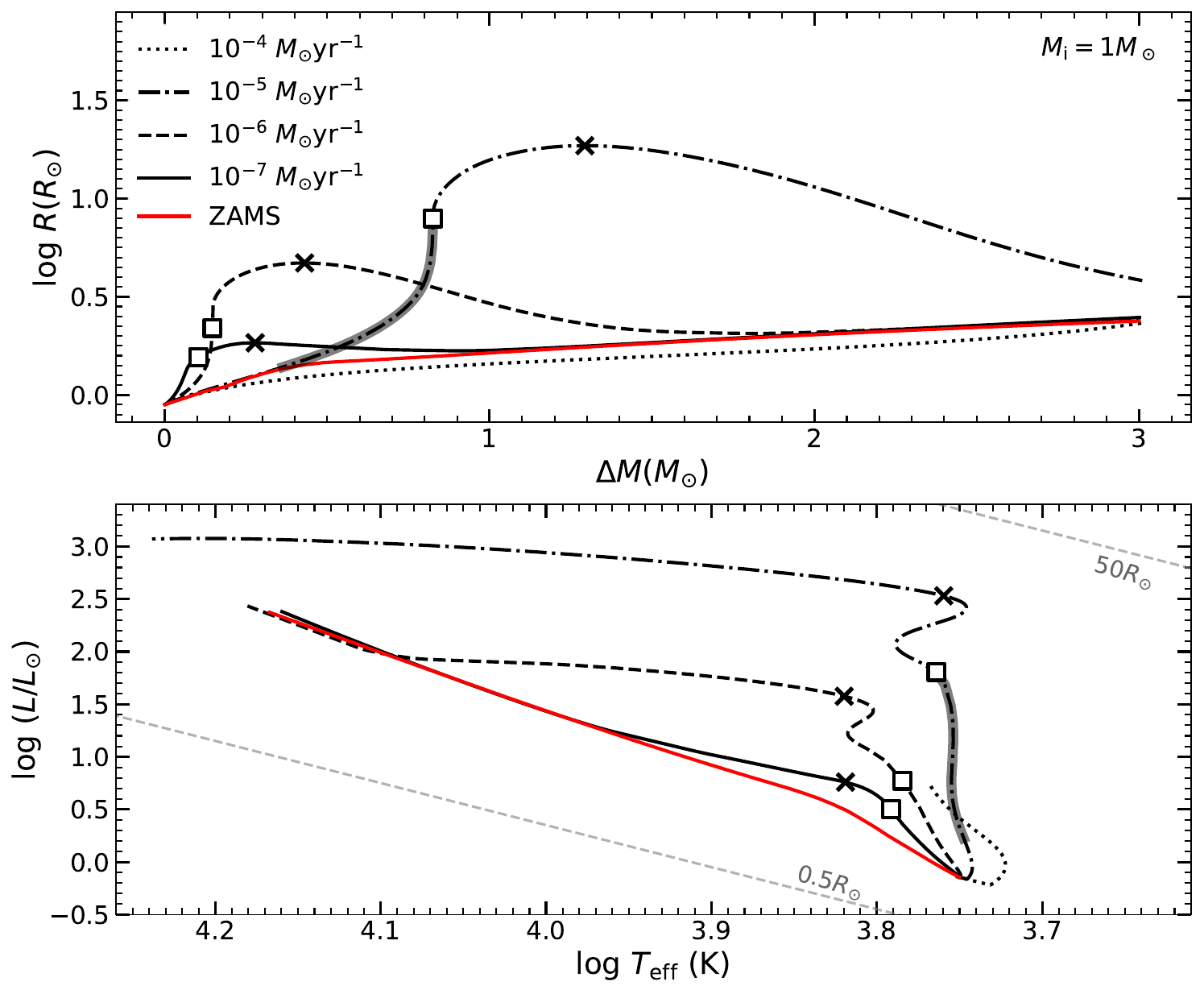}

  \end{minipage}%
   \caption{\label{fig2}: \textbf{Upper panel:} The radial variations of accretors with initial masses of 1$M_{\odot}$. The adopted accretion rates are shown in the legend. The open squares represent the disappearance of the convection envelope. The thick grey line indicates the stage where the convective envelope gradually decreases (only the case of $10^{-5}M_\odot\;\rm yr^{-1}$ is shown for clarity). The crosses represent the stage of the maximum radius. The red solid line represents the ZAMS radius for accretors with given masses. The thermal timescale mass transfer rate for a $1M_\odot$ accretor is about $2.7\times 10^{-8}M_\odot\;\rm yr^{-1}$.
   \textbf{Lower panel:} Evolutionary tracks of accretors with initial masses of 1$M_{\odot}$.
   }
\end{figure}

Although there is a shallow convective envelope on a star's surface, it will significantly affect the radius variation \citep{2001ApJ...555..990B,2005essp.book.....S}. Here, we use an example of $1M_{\odot}$ accretor to investigate the radius variation of stars with shallow convective envelopes on the surface, where the mass of the convective envelope initially is about 0.03 $M_{\odot}$. In the upper panel of Figure~\ref{fig2}, we present the relationship between the accreted mass and radius. The corresponding evolutionary tracks of the accretor are shown in the lower panel of Figure~\ref{fig2}. The mass transfer rates are given from $10^{-7} M_{\odot}\; \rm{yr^{-1}}$ to $10^{-4} M_{\odot}\;\rm{yr^{-1}}$, which are greater than the $1M_{\odot}$ accretor thermal timescale mass transfer rate of $2.70{\times}10^{-8} M_{\odot}\; \rm{yr^{-1}}$. The radius is larger than the ZAMS radius when $1M_{\odot}$ accretor accumulates mass, and the evolutionary track moves away from ZAMS in the HR diagram, which is similar to the $5M_{\odot}$ accretor.

We take the accretor with $\dot{M}$ =  $10^{-5} M_{\odot}\;\rm{yr^{-1}}$ as an example. In the beginning, the accretor has a non-negligible convective envelope. In the convective zone, the entropy gradient vanishes so that compression no longer releases heat, and almost all of the heat flux from the inner radiative shell is absorbed by the convective zone \citep{1989ApJ...341..306F}. Therefore, the radius expands slowly and follows the ZAMS radius line. With the increase of star mass, the internal temperature of the star rises and the energy transfer is gradually dominated by radiation, as shown in a thick grey line, where the open square represents the complete disappearance of the convective envelope.
The accretor then expands until the maximum radius is reached, similar to the $5 M_{\odot}$ accretor shown in Fig~\ref{fig1}. 
For the case of $10^{-4} M_{\odot}\; \rm{yr^{-1}}$, the time elapsed during the accretion processes is only 30000 years. In such a short timescale, the internal structure of the star does not have enough time to adjust itself, and the convective envelope exists during the whole accretion process. Therefore, the radius is always smaller than the corresponding ZAMS radius, as shown in the upper panel.

\subsubsection{Accretor with deep/fully convective envelope}
\label{result4}

\begin{figure}
  \begin{minipage}[htp]{0.45\textwidth}
  \centering
   \includegraphics[scale=0.33]{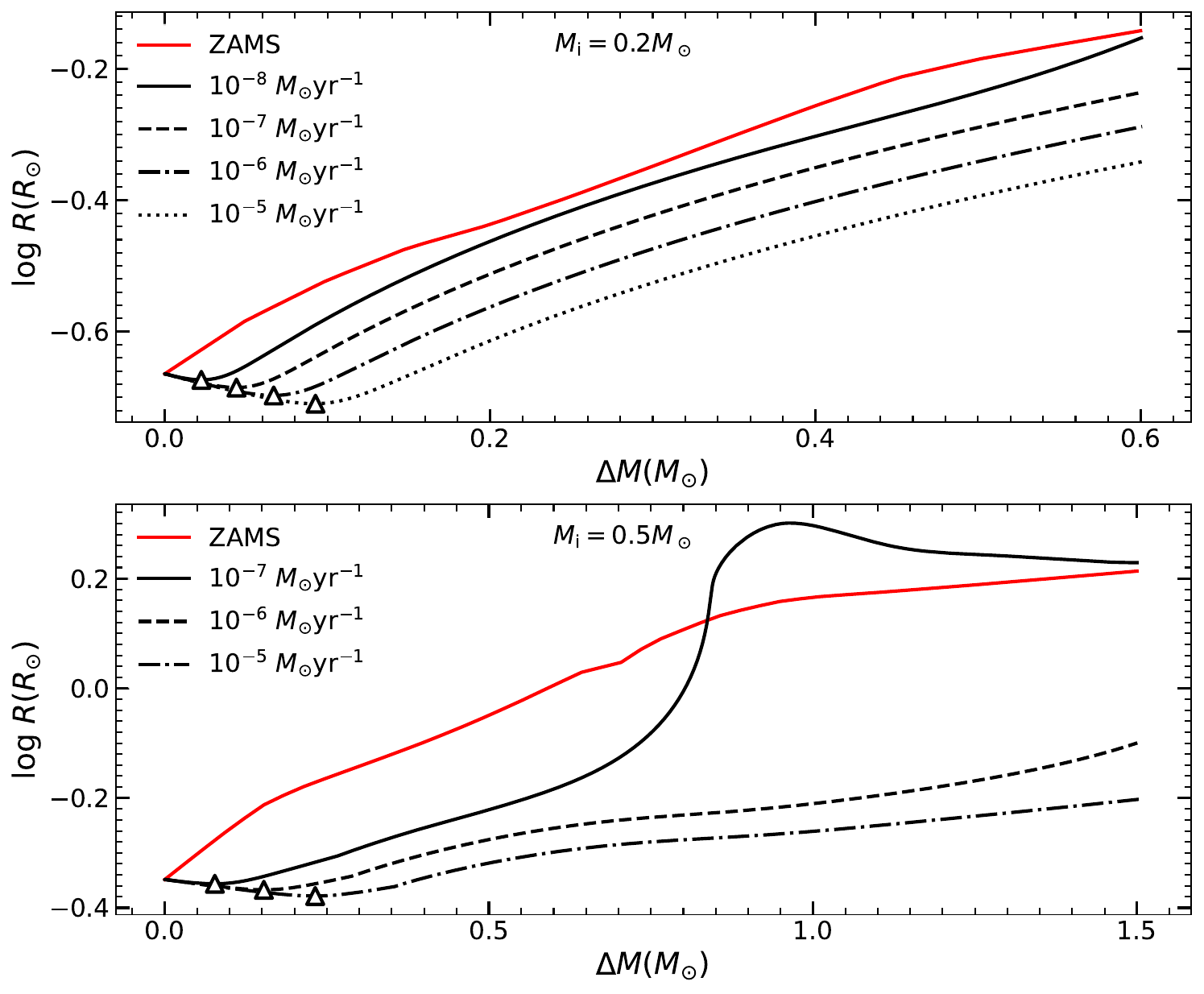}

  \end{minipage}%
  \caption{\label{fig3} The radial variations of accretors with initial masses of 0.2$M_{\odot}$ (upper panel) and 0.5$M_{\odot}$ (lower panel). The open triangles represent the stage of the minimum radius. The adopted accretion rates are shown in the legends. The red solid line represents the ZAMS radius for accretors with given masses. The thermal timescale mass transfer rate for accretors with masses of $0.5M_\odot$ and $0.2M_\odot$ are $1.5\times 10^{-9}M_\odot\;\rm yr^{-1}$ and $3.26\times 10^{-10}M_\odot\;\rm yr^{-1}$, respectively.
  }
\end{figure}

This section explores the radius changes for a deep convective star of $0.5M_\odot$ and a fully convective star of $0.2M_\odot$ with mass accumulation.
The thermal timescale mass transfer rates are about $1.50{\times}10^{-9}$ $M_{\odot}\; \rm{yr^{-1}}$ for a $0.5M_\odot$ accretor and $3.26{\times}10^{-10}$ $M_{\odot}\; \rm{yr^{-1}}$ for a $0.2M_\odot$ accretor, which are all smaller than the given mass transfer rates. Figure~\ref{fig3} shows the radius variations of the accretors with an initial mass of 0.2$M_{\odot}$ (the upper panel) and 0.5$M_{\odot}$ (the lower panel), and the HR diagrams of these accretors are addressed in Appendix \ref{appendix}. In the case of deep/fully convective stars, the radius initially decreases since the accretion energy is transferred
to the star interior by convection \citep{1968Natur.220..143I,2015MNRAS.453..666C}. The effective temperature then increases with the contraction of the star. The stars become denser and hotter, generating more nuclear energy, leading to the radial expansion in the subsequent evolution. 

For the case of a $0.5M_\odot$ accretor with the mass transfer rate of $10^{-7} M_{\odot}\; \rm{yr^{-1}}$, as shown in the lower panel of Figure~\ref{fig3}, the radius rapidly expands until it reaches the minimum value. After the convective envelope of the star disappears, the accretor expands beyond its corresponding ZAMS and then returns. For the higher mass transfer rates of $10^{-6} M_{\odot}\; \rm{yr^{-1}}$ and $10^{-5} M_{\odot}\; \rm{yr^{-1}}$, the convective envelope always exists, resulting in a radius smaller than the corresponding ZAMS radius. These accretors are less likely to fill their Roche lobe and enter the CE phase.

\subsection{Dynamical mass transfer stability criterion}
\label{result5}

The mass transfer stability is one of the most fundamental questions in binary evolution. In general, the stability criterion of the binary interaction can be understood via the donor response due to the mass loss \citep{2013A&ARv..21...59I}. In an extreme case, if the donor departs from the hydrostatic equilibrium with mass loss, the mass transfer will be dynamic. The binary may enter the CE phase (see more details in \citealt{2010ApJ...717..724G,2020ApJS..249....9G}). However, as we discussed above, the accretion may also lead to the radial expansion of the accretor. In this section, we investigate the mass transfer stability criterion via the response of the accretor due to the mass accumulation. We assume that if the radius of the accretor is greater than the outer Lagrangian radius, the unbound material will engulf the binary, and resulting in the CE phase \citep{2004ApJ...601.1058I,2020ApJ...899..132G,2023MNRAS.519.1409L,2023A&A...669A..45T}. 

The dimensionless radius of the outer critical surface is given by \citet{2020ApJS..249....9G}, approximate analytic fits to integrations of the Roche limit by \citet{1984ApJS...55..551M} and \citet{1985ibs..book..197P}
\begin{equation}\label{eq5}
\begin{split}
r_{\rm L_2}(q) = \frac{R_{\rm L_2}}{A} = r_{\rm L}(q) +[0.179+0.01(\frac{q}{1+q})]\\ \times(\frac{q}{1+q})^{0.625}\quad{\text{for}~{q\leq 1}}
\end{split}
\end{equation}  
or
\begin{equation}\label{eq6}
\begin{split}
r_{\rm L_3}(q) = \frac{R_{\rm L_3}}{A} = r_{\rm L}+[0.179+0.01(\frac{q}{1+q})- 0.025(\frac{q-1}{q})]\\
\times(\frac{q}{1+q})^{0.625} q^{-0.74} \quad{\text{for}~{q\geq 1}}.
\end{split}
\end{equation}  
where $R_{\rm L_{2/3}}$ is the volume-equivalent radius of the donor's outer lobe. $r_{\rm L}(q) = {0.49q^{\frac{2}{3}}}/({0.6q^{\frac{2}{3}} + \text{ln}(1+q^{\frac{1}{3}})})$. A is the binary distance, and $q$ is the mass ratio. $q = M_\text{1}/M_\text{2}$, $M_\text{1}$ is the donor and $M_\text{2}$ is the accretor. And for the accretor, we need to change $q$ to $1/q$.  

\begin{figure}
 \begin{minipage}[t]{0.45\textwidth}
  \centering
   \includegraphics[scale=0.34]{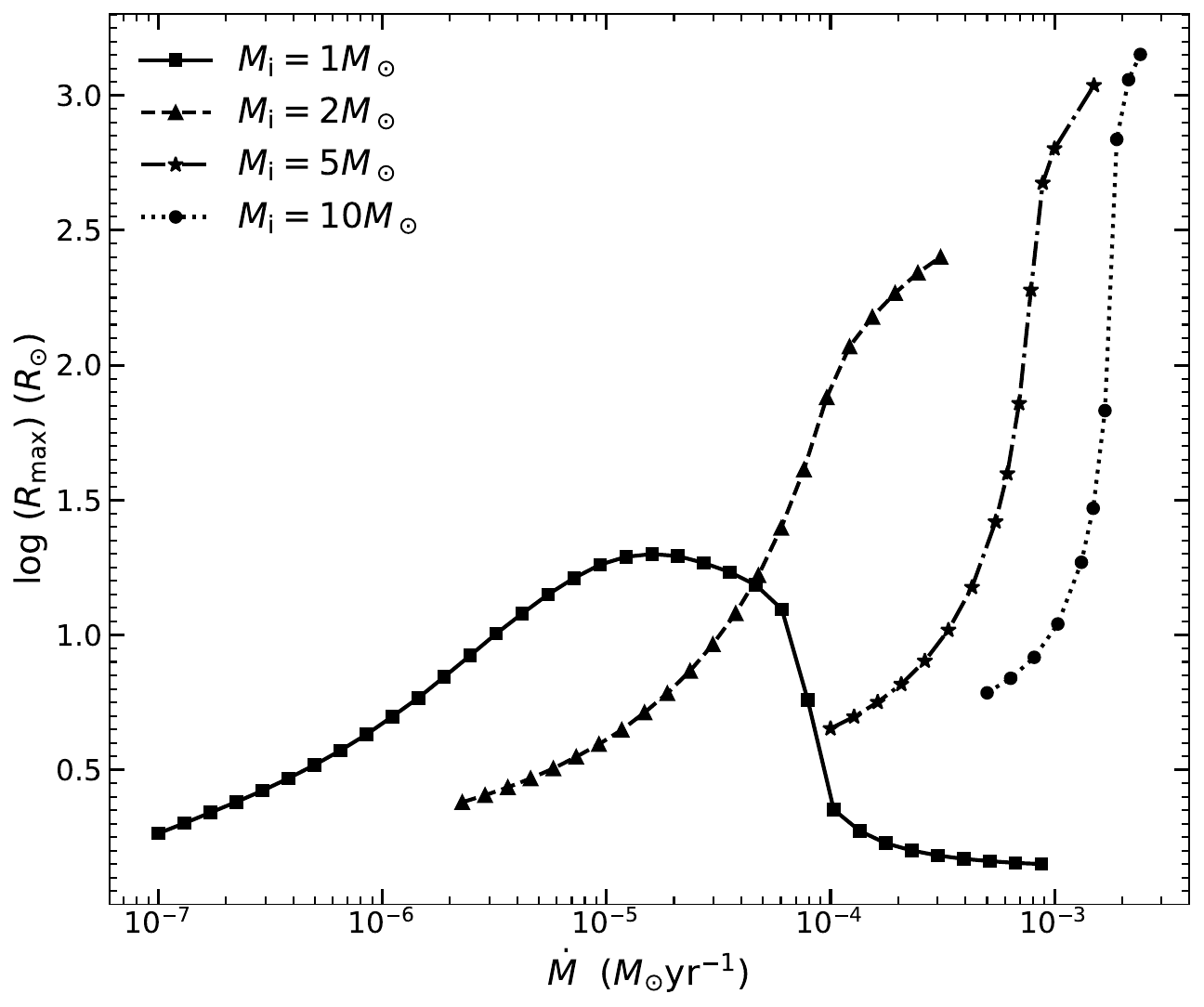} 

  \end{minipage}%
  \caption{\label{fig5} The relationship between maximum radius and mass transfer rate for accretors with initial masses of 1$M_{\odot}$, 2$M_{\odot}$, 5$M_{\odot}$, and 10$M_{\odot}$. Each symbol for a given line represents a distinct MESA grid in our simulations, with line style indicating the initial accretor mass.}
\end{figure}

At first, we calculate the maximum radius of the accretor with mass accumulation. For accretors with deep-convection envelopes, the radial expansion is small or the stars shrink during the accretion processes, which are not considered here for clarity. 
Four typical examples for accretors with masses of $1M_\odot, 2M_\odot, 5M_\odot$ and $10M_\odot$ based on the above results as investigated. In Figure~\ref{fig5}, we show the maximum radii of these accretors with different mass transfer rates. For stars with masses of 2$M_{\odot}$, 5$M_{\odot}$, and 10$M_{\odot}$, the mass transfer rates increase, leading to a larger maximum radius. However, the maximum radius of 1$M_{\odot}$ accretor decreases when the mass transfer rate exceeds $\sim 10^{-4} M_{\odot}\; \rm{yr^{-1}}$, since the convective envelope exists during the accretion process, inhibiting further radius variation, as discussed in Section \ref{result3}.

For a real mass transfer process, the mass transfer rate changes with evolutionary age. \citet{1977A&A....54..539K} proposed that the mean mass exchange rate can be a good approximation for studying binary evolution, where the mean mass transfer rate represents the time interval from the beginning of mass exchange to the loss of half of the original mass of the donor. 
In this work, we performed several binary evolution simulations via MESA and found a fitted formula between the mean mass transfer rate and the initial donor mass for the case of initial mass ratio, $q_{\rm i}=1$, i.e., 
The fitted formula is given
\begin{equation}\label{eq7}
\begin{split}
\log \dot{M} = -4.27\times (\log M_{\rm 1i})^2 +11.70\times \log M_{\rm 1i} - 10.57,
\end{split}
\end{equation}
where $M_\text{1i}$ is the mass of the donor, in units of $M_\odot$, and $\dot{M}$ is the mean mass exchange rate, in units of $M_{\odot}\;\rm{yr^{-1}}$. The detailed inputs of binary simulations are given in Appendix  \ref{appendixB}. It should be noted that for donor stars more massive than $\sim 25M_\odot$, the wind mass loss rate is so large that the approximation of the mean mass transfer rate is unsuitable. Therefore, the fitted formula of Eq. \ref{eq7} only works for the case of $M_{\rm 1i}\lesssim 25M_{\odot}$. With the increase of the initial mass ratio, the values of mean mass transfer rates will slightly increase, as shown in Figure~\ref{fig16} of Appendix  \ref{appendixB}. Its influence on our results is discussed below. 

\begin{figure}
  \begin{minipage}[t]{0.495\linewidth}
  \centering
   \includegraphics[scale=0.34]{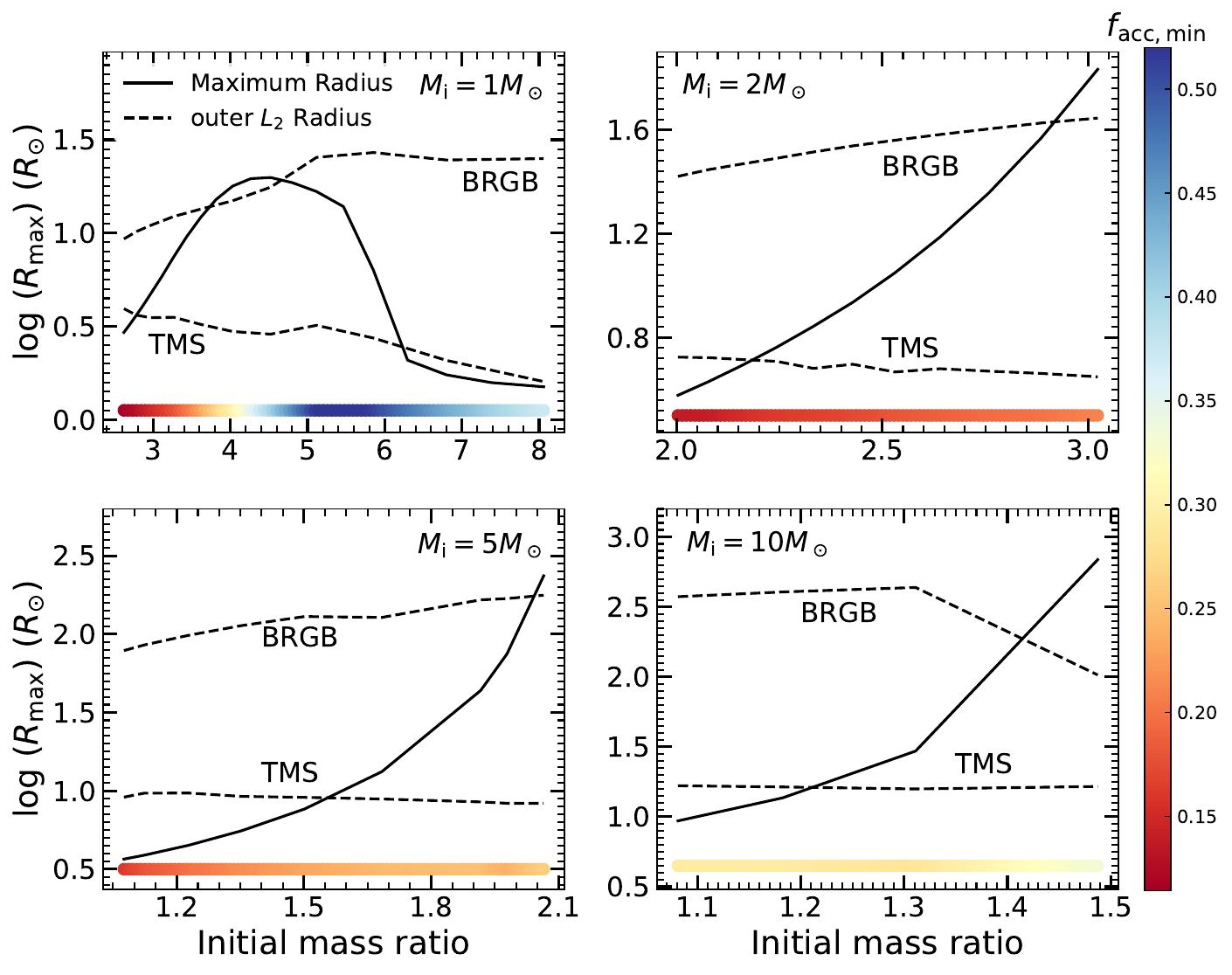}

  \end{minipage}%
  \caption{\label{fig6} The maximum radius is a function of the initial mass ratio for different initial accretors. The corresponding outer Roche lobe radii after the mass transfer processes for the donor star filling its Roche lobe at different evolutionary stages are shown in dashed lines, where TMS means the termination of MS and BRGB means the base of RGB.  If the solid line is above one of the dashed lines, it means the maximum radius for a given initial mass ratio is larger than the corresponding outer $L_2$ radius and the binary would enter into the CE phase.The colors show the minimum accretion efficiency $f_{\rm acc,min}$. If the mass transfer efficiency is larger than $f_{\rm acc,min}$, the accretor can accumulate enough matter and make the radius increase to the maximum value. }
\end{figure}

For a given accretor and the mass transfer rate, we could know the donor mass. Then, we convert the values of the mass transfer rate to the mass ratio for a given accretor. The maximum radius of the accretor as a function of $q_{\rm i}$ is shown in Figure~\ref{fig6}, where the initial accretor masses are labeled in each panel. 
For a star filling its Roche lobe at an evolutionary stage, one could obtain the initial orbital separation at the onset of mass transfer. Assuming that mass transfer is fully conserved, we further calculated the orbital separation when $\Delta M$ has been accreted by the accretor based on angular momentum conservation, where $\Delta M$ is the accreted mass as the accretor reaches its maximum radius. Then we get the corresponding outer Roche lobe radii of the accretors accoding to Equations \ref{eq5} and \ref{eq6}, as shown in the dashed lines of Figure~\ref{fig6}. If the maximum radius (solid line) is larger than the outer Roche lobe radius at a certain evolutionary stage, the binary will experience the dynamically unstable mass transfer. The cross points between the solid lines and the dashed lines then define the critical mass ratios for donors at certain evolutionary stages, above which the binary would enter the CE phase. We see that for accretors with masses of $1 M_\odot, 2 M_\odot$ and 5$ M_\odot$, the critical mass ratios for donors filling the Roche lobe at the termination of MS (TMS) stages are about $\sim 1.5-2.8$ (The corresponding cross points between the solid line and the TMS line), and for the 10$ M_\odot$ accretor is about 1.2, which are lower than critical mass ratios derived from the donor response (about $3$; e.g. \citealt{2002MNRAS.329..897H,2020RAA....20..161H,2020ApJS..249....9G}). For stars beginning to transfer material at Hertzsprung gap (HG) stages (between the TMS and the base of the red giant branch (BRGB) lines), the critical mass ratio derived from the donor response is about $4$, which is still larger than the value derived based on the accretor response for the cases of $M_{\rm 2i}\gtrsim 2M_\odot$. 
Our results may suggest that the binary is easier to enter the CE phase for a donor star at the MS or HG stage than previously believed. 

For a real binary mass transfer process, the accretor may not be able to accrete all materials. In order to illustrate the influence of mass transfer efficiency on our results, we calculate the values of minimum accretion efficiency, $f_{\rm acc,min}$, where $f_{\rm acc,min} =  \Delta M_2/\Delta M_1$, where $\Delta M_2$ is the accreted mass as the accretor reaches its maximum radius and $\Delta M_1 \equiv M_{\rm 1i}$ is the transferred mass (we simply assumed that all of the donor masses are lost\footnote{If the donor star has a developed core, not all material can be lost. In this case, we found the values of $f_{\rm acc,min}$ increase no more than $\sim 10\%$ for the core mass fraction of less than $\sim 0.3$.}). If the mass transfer efficiency is larger than $f_{\rm acc,min}$, the accretor can accumulate enough matter and make the radius increase to the maximum value. The values of $f_{\rm acc,min}$ are given in Figure~\ref{fig6}. For the case of $1M_\odot$ accretor, $f_{\rm acc,min}$ ranges from $\sim 0.15$ to $0.5$, and for the cases of accretor masses $\gtrsim 2M_\odot$, $f_{\rm acc,min}$ ranges from $\sim 0.15$ to $0.3$. The specific value of mass transfer efficiency for a binary is still unclear. In the recent work, \citet{2021ApJ...908...67S} argued that a large accretion efficiency ($\sim 0.5$) seems to match Be binaries in the observations better. In this case, our conclusions on the critical mass ratio are generally valid. Besides, our results are based on the fitted formula of mean mass transfer and initial donor mass, i.e., Eq. \ref{eq7}, where the initial mass ratio is set to be $1$. For a large value of the initial mass ratio, we found the mean mass transfer rate slightly increased for a given initial donor star, as shown in Figure~\ref{fig16}. The large mean mass transfer rate generally leads to a larger value of maximum radius. Therefore, the binary will more easily enter into the CE phase, resulting in the smaller critical mass ratios derived above. Our estimation of the mass transfer instability can be improved and extended with further detailed binary evolution simulations.

\section{Summary and Conclusion}
\label{sect:conclusion}
In this letter, we investigate the evolution of accretors with mass accumulation. Several cases are considered, including stars with a radiative envelope, shallow convective envelope, deep convective envelope and fully convective stars. Assuming that the binary would enter into the CE phase if the radius of accretors exceeds the outer Roche lobe radius, we obtain the critical mass ratio of the dynamically unstable mass transfer stability criterion based on the accretor response. Our main conclusions are summarized as follows.

(1) For stars with a radiative envelope, the radius will rapidly increase during accretion until it reaches its maximum. The radius decreases when the mass transfer is smaller than the thermal timescale mass transfer, and the star returns to the ZAMS.

(2) For stars with a convective envelope, the convection will suppress the radius change. The radius is relatively stable before the convection disappears. For a star with a deep convective envelope or a fully convective star, the radius is smaller than that of ZAMS with the same mass.

(3) We find the critical mass ratios for donors filling their Roche lobes at MS and HG stages are smaller than that derived from the donor response. Our results may suggest that the binary is easier to enter into the CE phase for a donor star at the MS or HG stage than previously believed. 

The results show that the accretor response is important in binary evolution simulations, and it may lead to incorrect conclusions by assuming a point mass for the accretor. It should be noted our results are obtained according to the constant accretion rate, which is a simplified approximation of the binary interaction. We plan to investigate the accretor response with detailed binary evolution simulations in the further study.

\section*{Acknowledgements}
We cordially thank the anonymous referee for the detailed comments that improved this manuscript. This work is supported by the National Key R$\&$D Program of China (Grant Nos. 2021YFA1600403, 2021YFA1600400), the Natural Science Foundation of China (Grant Nos. 12125303, 12090040/3, 12103086, 11521303, 11703081, 11422324, 12288102), the Natural Science Foundation of Yunnan Province (No. 202201BC070003), the Yunnan Revitalization Talent Support Program—Science $\&$ Technology Champion Project (No. 202305AB350003), by the National Ten-thousand Talents Program, by Yunnan Province (No. 2017HC018), the Youth Innovation Promotion Association of the Chinese Academy of Sciences (Grant no. 2018076), the Key Research Program of Frontier Sciences of CAS (No. ZDBS-LY-7005), Yunnan Fundamental Research Projects (Grant Nos. 202301AT070314, 202101AU070276, 202101AV070001) and the International Centre of Supernovae, Yunnan Key Laboratory (No. 202302AN360001). We also acknowledge the science research grant from the China Manned Space Project with No. CMS-CSST-2021-A10.
Software: MESA (12115; \citealt{2011ApJS..192....3P,2013ApJS..208....4P,2015ApJS..220...15P,2018ApJS..234...34P,2019ApJS..243...10P}).

\section*{Data Availability}

The data underlying this article will be shared on reasonable request to the corresponding author.



\bibliographystyle{mnras}
\bibliography{bibtex} 




\appendix

\section{Evolutionary tracks for $0.2M_{\odot}$, $0.5M_{\odot}$, $2M_{\odot}$ and $10M_{\odot}$ accretors }
\label{appendix}

\begin{figure}
  \begin{minipage}[h]{0.495\textwidth}
  \centering
   \includegraphics[scale=0.33]{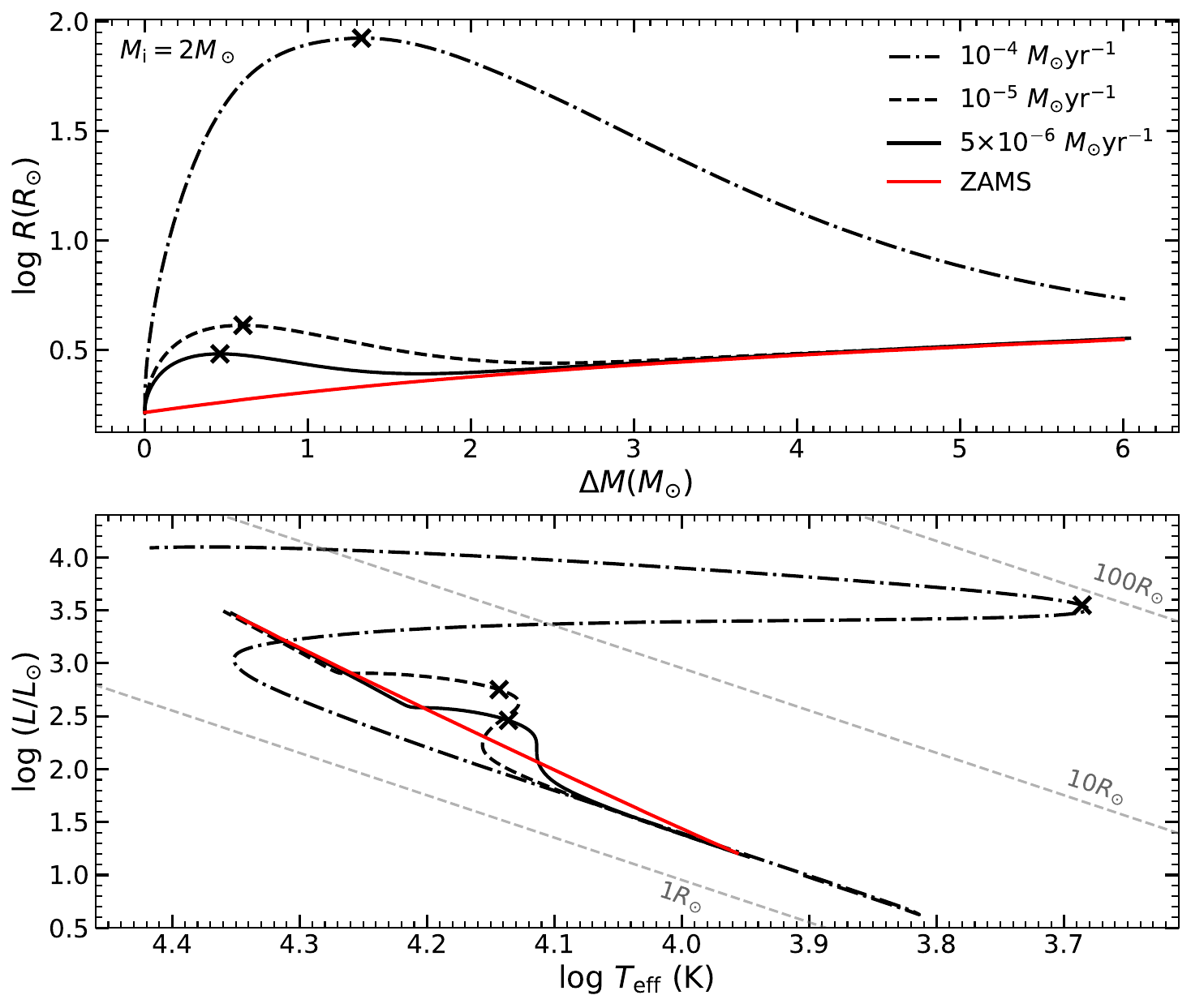}
   
  \end{minipage}%
\caption{\label{fig12} Similar to Figure~\ref{fig1} but for the $2M_\odot$ accretor. }  
\end{figure}

\begin{figure}
  \begin{minipage}[h]{0.495\textwidth}
  \centering
   \includegraphics[scale=0.33]{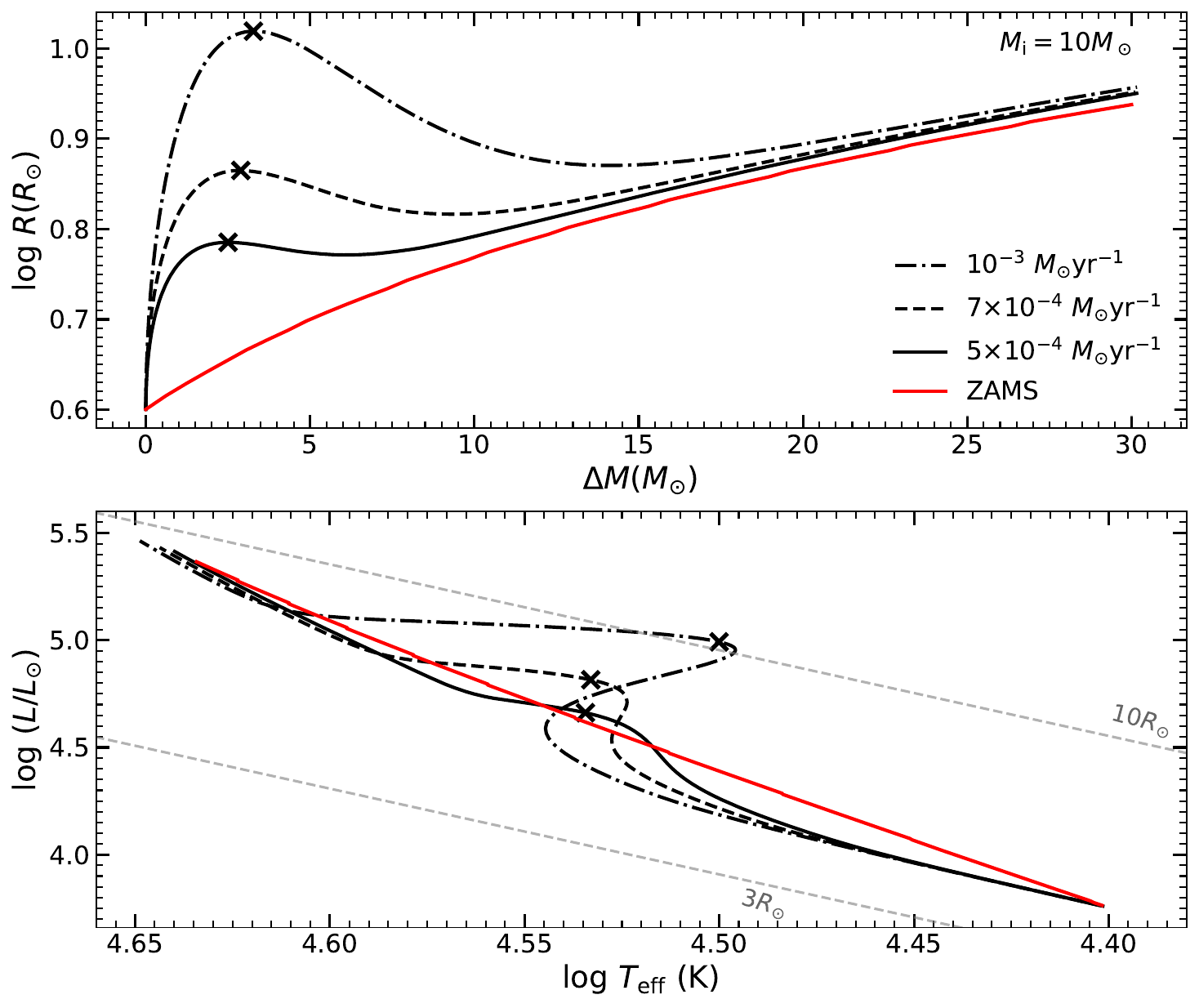}

  \end{minipage}%
  \caption{\label{fig13} Similar to Figure~\ref{fig1} but for the $10M_\odot$ accretor.  }
\end{figure}

\begin{figure}
  \begin{minipage}[h]{0.495\textwidth}
  \centering
   \includegraphics[scale=0.38]{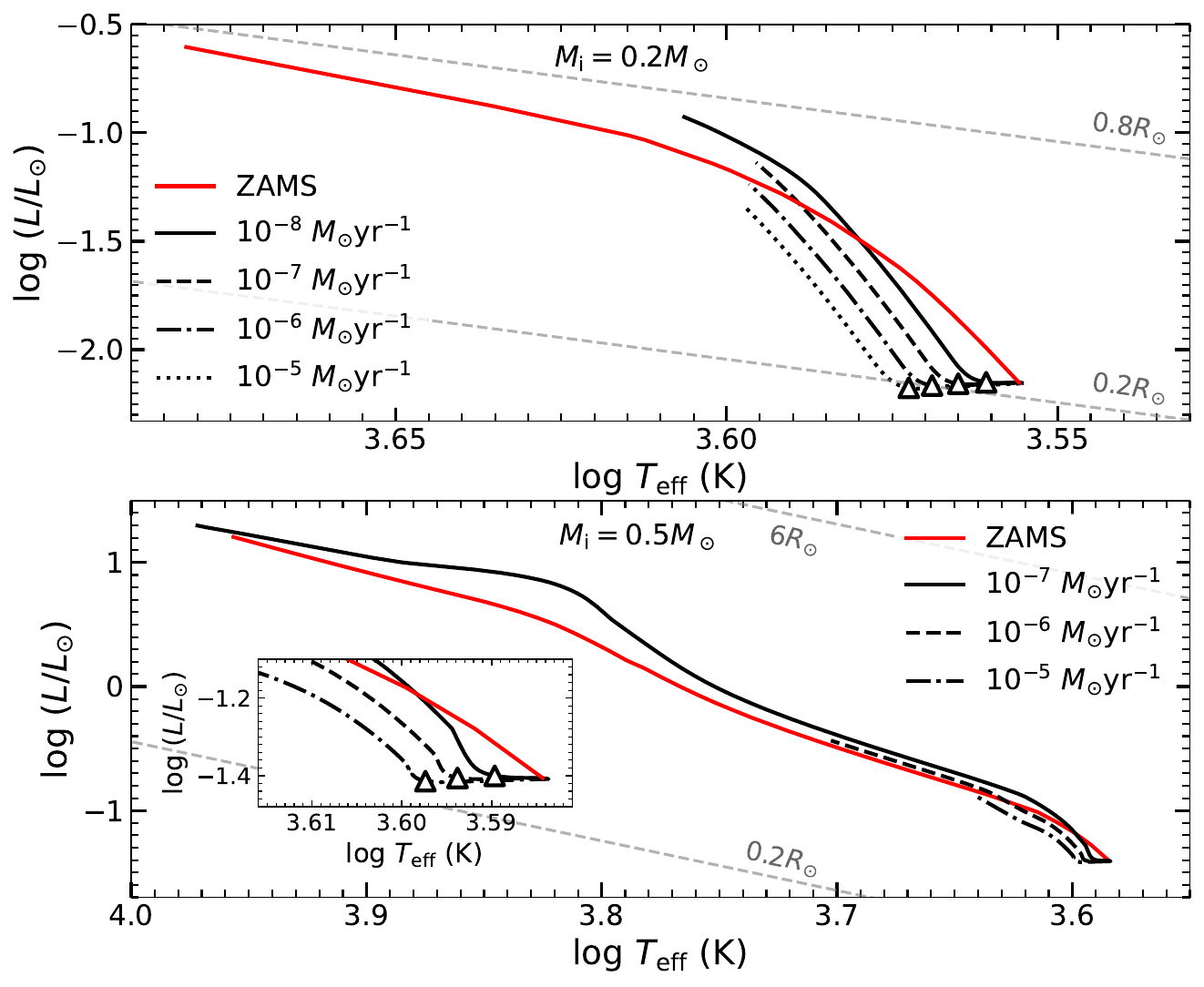}

  \end{minipage}%
  \caption{\label{fig14}  Evolutionary tracks of accretors with 0.2$M_{\odot}$ and 0.5$M_{\odot}$. The open triangles represent the stage of the minimum radius. The meanings of other symbols are the same as in Figure~\ref{fig1}.  }
\end{figure}

Figure~\ref{fig12} shows the radius variation and HR diagram of the $2M_{\odot}$ accretor. And the $10M_{\odot}$ accretor radius variation and HR diagram are represented in Figure~\ref{fig13}. Figure~\ref{fig14} shows the evolutionary tracks of accretors with 0.2$M_{\odot}$ and 0.5$M_{\odot}$.

\section{Mean mass transfer rate}
\label{appendixB}
\begin{figure}
  \begin{minipage}[t]{0.495\linewidth}
  \centering
   \includegraphics[scale=0.34]{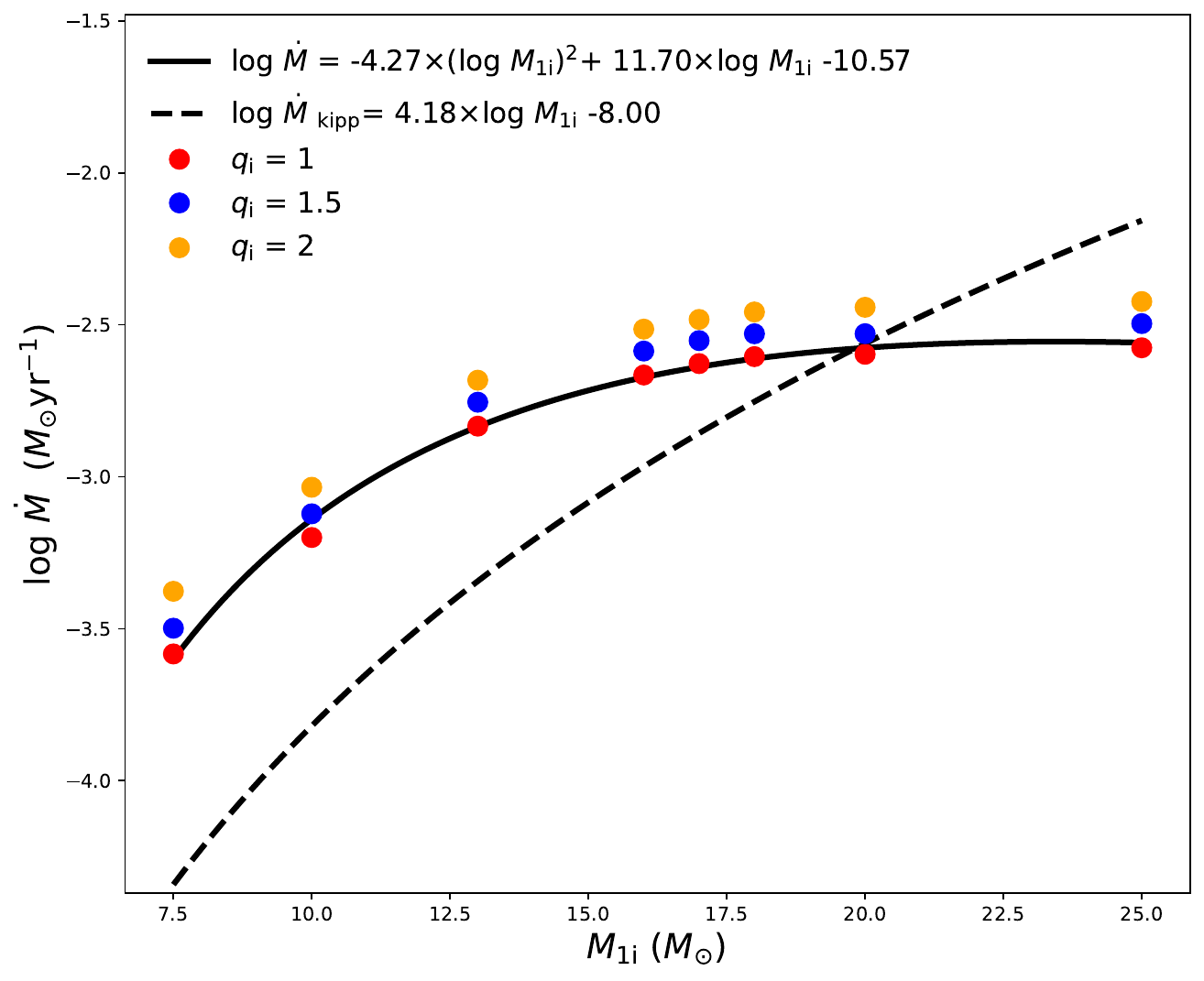}

  \end{minipage}%
  \caption{\label{fig16} The relationship between the mean mass transfer rate and the donor mass. The corresponding colored dots represent different $q_{\rm i}$. The black solid line represents the fitted curve based on the simulated results of $q_{\rm i}=1$, and the black dashed line is taken from \citet{1977A&A....54..539K}.}
\end{figure}

We evolve several binary systems with the MESA binary module. The initial donor masses are set to be $7.5$, $10$, $13$, $16$, $17$, $18$, $20$ and $25$, and three values of initial mass ratio, i.e., $q_{\rm i}=1,1.5,2$, are considered. For each binary, the mass transfer occurs approximately at the early HG gap. The ``Kolb'' mass transfer scheme is chosen \citep{1990A&A...236..385K}. The main inputs, such as stellar wind and angular momentum loss, are adopted following \citet{2022A&A...659A..98S}. Figure~\ref{fig16} shows the relation between the mean mass transfer rate and the donor mass. The simulation results are given in colored circles and the solid line is for the fitted curve of Eq. \ref{eq7}. The results from \citet{1977A&A....54..539K} are also given for comparison.

\bsp	
\label{lastpage}
\end{document}